\documentclass[twocolumn,floats,prl,aps]{revtex4}

\usepackage{graphicx}

\newcommand\beq{\begin{equation}}
\newcommand\eeq{\end{equation}}
\newcommand\bea{\begin{eqnarray}}
\newcommand\eea{\end{eqnarray}}
\newcommand{\nonum}{\nonumber}

\begin{document}

\title{\bf Adiabatic Cooper-Pair Pumping in a Linear Array of   
Cooper Pair Boxes}

\author{\bf Sujit Sarkar}
\address{\it 1. PoornaPrajna Institute of Scientific Research,
4 Sadashivanagar, Bangalore 5600 80, India\\
}

\date{\today}

\begin{abstract}
We present a study of adiabatic Cooper pair pumping in one dimensional
array of Cooper pair boxes. We do a detailed theoretical analysis
of an experimentally 
realizable 
stabilized charge pumping scheme in a 
linear array of Cooper pair boxes. Our system is subjected to
synchronized flux and voltage fields and travel along a loop which
encloses their critical ground state of the system in the flux-voltage plane. 
The locking potential in the sine-Gordon model slides and changes its minimum
which yields the Cooper pair pumping. Our analytical methods are the 
Berry phase analysis and Abelian bosonization studies. 

\end{abstract}

\maketitle


{\bf 1. Introduction:}
The adiabatic pumping physics gets more attraction after the
pioneering work of Thouless \cite{thou1,thou2}. 
Quantum adiabatic pumping physics is related with many systems 
like open quantum dots \cite{brou,alt,dot},
superconducting quantum wires \cite{bla, wire}, Josephson junctions \cite{pek,levy}
, the Luttinger quantum wire \cite{lutt1}, 
interacting
quantum wire \cite{lutt2} and also to the quantum spin pump \cite{shin}.\\
An adiabatic parametric quantum pumping is a device that generates a
dc current by a cyclic variation of system parameters, the variation 
being slow enough that the system remains close to the ground state throughout
the pumping cycle. It is well known to us that when a quantum mechanical system
evolves then it acquires a time dependent dynamical phase and time independent
geometrical phase \cite{berry}. 
Geometrical phase dependent on the geometry of the path
turned in the parameter space. For the closed path encircling the critical
ground state differ by the phase factor and the cyclic parametric variation
introduce non-vanishing transport in the system. 
Quantum pumps are to transform 
an ac signal of frequency $f$ into dc current given by the relation $I =f Q$,
where $Q= ne$ for electron pumping and $Q = 2 ne$ for Cooper pair pumping.
Quantum pumping with perfect accuracy could be utilized to establish a
standard of current. Here we have considered the perfect pumping condition.
The error in the pumping procedure arises due to the current reversal and
the spontaneous charge excitations. In Ref. (\cite{ravi} ), the authors 
have discussed the sources
of errors (the non-adiabatic correction leave the system
in an unknown superposition of the charge state, instead of 
definite charge state, after the full cycles)  and their minimization.
Here we would like to explain the theoretical detailed of
that experimental proposals stabilized charge pumping \cite{ravi} in a linear
array of Cooper pair boxes. 
Before
we start our full swing quantum field theoretical calculations. We would like  
to present a derivation to illustrate the Cooper pair pumping in one
dimensional array of Cooper pair boxes. 
 
\section{ 2. Model Hamiltonians and Continuum Field Theoretical Study:}

Our Hamiltonian ($ H= {H_c} + {H_J}$) 
consists of two parts one is the Josephson coupling and the
other is the Coulomb charging energy. Where
${H_J}= -\sum_{k=1}^{N} {E_{J,k}} cos {{\phi}_k} $
and $ {H_c} = \sum_{k=1}^{N} \frac{{Q_k}^2}{2 C_k} $. $Q_k$ is
the charge occurs at the kth junction, this charge is measured
w.r.t the gate charge. So the effective charge of the k'th junction can
be varied by varying the gate voltage of the system connected to
the junctions. The charging Hamiltonian is diagonal in the basis 
formed by the charge eigen state 
$ |\vec{n} > = | {n_1}, {n_2},.......{n_{N-1}} >$, $ n_i $ is the number
of Cooper pair in each island. We would like to explain the transport of
Cooper pair of this one dimensional tunnel junction through the analysis
of Berry phase \cite{berry}. 
We will see that the Cooper pair current in the system consists 
of two parts
one is the conventional supercurrent and the other is adiabatic Cooper
pair pumping. Here we follow the seminal paper of Berry \cite{berry} for our theoretical
analysis. In our system, we are varying the Josephson junction couplings
through the applied flux and the charge on the dot by applying the gate
voltage. So the adiabatic varying parameters are, 
$R = ({E_J}, {Q_k})$. The state $| {\psi} (t) >$, of the
systems evolves according to Schrodinger's equation
$$ H (R (t) ) | {\psi (t)} > = i h  |{\psi (t)} > .$$ At any instant, the
natural basis consists of the eigenstates 
$|n(R) > = |{n_{1R}}, n_{2R},..... {n_{ N-1R}}>$. 
$|{\psi (t_0 )} > = | n (t_0 ) >$.
 
\bea
|{ \psi (t) }> & = & e^{ -i {E_n} \frac{\delta {t}}{h} } |n (t_0 ) >  
+ \sum_{l \neq n} \frac{h}{i} \frac{F({E_l}, {E_n})}{( E_{l0} - E_{m0} )} 
  \times \nonum\\
 & & <l (t_0 )| {{\nabla}_R} n (t_0 ) >. {{\delta}_t} R | l (t_0 ) >  
\eea
where $ F({E_l}, {E_n}) = ( e^{-i {E_l} ({t_0}) } - e^{-i {E_n} ({t_0}) } )$
    
\beq
|\psi (t) > =  |n (t) > ~+~ |{\delta {n}} (t) >
\eeq
${\nabla}_ {R}$ is the derivative w.r.t externally
varied parameter. The amount of charge that passes through the 
tunnel junction k during $\delta {t}$ interval of time is 
\beq
\delta {Q_k} ~=~ \int_{t_0}^{t} < {\psi} (t) | {I_{S,k}} | {\psi} (t) > dt
\eeq 
\beq
\delta {Q_k} ~=~ \delta {t} < I_{S,K} >_{n (t_0 )}
+ 2 Re [ \int_{t_0}^{t} <n (t)| I_{S,k} | {\delta n} (t) >].  
\eeq
Where 
\beq
I_{S,k} ~=~ \frac{-2 e}{2 h} \sum_{k}  E_{J,k}
(-i | {n_R} + {{\delta}_k} > <n_{R}| e^{i \frac{\phi}{N}} + h.c |) 
\eeq
${\delta}_k$ measure the change of $n$ due to the tunneling of one
Cooper pair through the kth junction.
The first term gives the direct Cooper pair transform, i.e., the supercurrent
. The second term corresponds to the adiabatic Cooper pair transform. The
charge transfer contribution for the second term is
\beq
\delta {Q_{1}} = -2 h \sum_{l \neq n} Im \frac{ [ <n| I_{S,k}|l> 
<l| {\delta} n >}{{E_l} - {E_n}} ] 
\eeq
The total charge transport due to Cooper pair transport over a
pumping period $\tau$ is  given
\beq
Q = \frac{-2e}{h} \int_{0}^{\tau} {\partial}_{\phi} {E_n} (t) dt
+ 2 h \sum_{l \neq n} Im [ \frac{ <n| I_{S,k}|l>
<l| {\delta} n >}{{E_l} - {E_n}} ]
\eeq
We are interested to build up the theory of experimental proposal of
Ref. \cite{ravi}. There system consists of N Cooper pair boxes. 
\begin{figure}
\includegraphics[scale=0.35,angle=0]{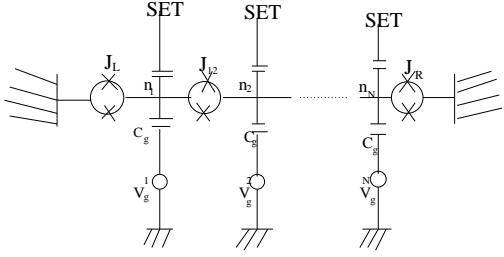}
\caption{The stabilized Cooper pair pumping system. Identical pulse
sequence is applied to every alternate superconducting island. SET
is the single electron transistors, they are biased with short voltage
pulses every time the pump should be in a definite charge state. Leftmost
and rightmost island are coupled to the reservoirs. ${n_i}$ is the
number of Cooper pair at the ith site. $V_g$ and $C_g$ are respectively
the gate voltage and capacitance of each Cooper pair box. Circle with two
crosses represents the Josephson junction.}
\label{Fig. 1}
\end{figure}
Nearest neighbor
are coupled with dc-SQUIDS. Leftmost and rightmost island are coupled
to the reservoirs  via other Squids as shown in Fig. 1. This figure
is nothing but the collection of tunable Josephson coupling and the
electrostatic potential of superconducting island in a 
Cooper pair box \cite{ravi,mak1,mak2}.
The adiabatic quantum Cooper pair pumping procedure,
is nothing but the transport of Cooper pair from one end of the system to
the other end.
The basic of Cooper pair tunneling in an array of Cooper pair boxes
can be understood from the analysis of Cooper pair transport in a
single Cooper pair box with two terminal Squid. One of the terminal
(say left) to transport a Cooper pair into the box and another to
transport the same pair to the reservoir. This process generate
the current.
The electrostatic energy of the
system can be expressed as 
\bea
E & = & \sum_{i} E_{c(i)} {( {n_i} ~-~{n_{g (i)}} )}^2 + \nonum\\
& & \sum_{i} E_{m (i,i+1)} {( {n_i} ~-~{n_{g (i)}} )}  
{( {n_{i+1}} ~-~{n_{g (i+1)}} )}.
\eea
Here $E_{c(i)}$ and $E_{m (i,i+1)}$ are respectively the charging energy
and the electrostatic couplings between two islands. We are interested
in the charge degeneracy point, i.e., when the gate charge is close to
1/2, the lowest energy states are characterized by either zero or one
Cooper pair on each island. With this assumption they have reduced the
Hilbert space and map the system to a finite anisotropic Heisenberg
spin-1/2 chain in an external magnetic field \cite{ravi}. 
In the spin language Cooper pair pumping is nothing but the transport of spin
(Jordan-Wigner fermions) from one
end of chain to the other end.
They have defined the
Hamiltonian \cite{ravi}
\bea
H & = & \frac{-1}{2} {{B_x}^{1}} {{\sigma}_x}^{1} - 
\frac{1}{2} {{B_x}^{N}}{{\sigma}_x}^{N} - 
\sum_{i}^{N} \frac{1}{2} {{B_z}^{i}}{{\sigma}_z}^{i} \nonum\\
& & + \frac{1}{2} 
\sum_{i=1}^{N-1} [ {\Delta}_{i,i+1} {{\sigma}_z}^{i}{{\sigma}_z}^{i+1} 
 -J_{i,i+1} ( {{\sigma}_{+}}^{i}{{\sigma}_{-}}^{i+1} +
{{\sigma}_{+}}^{i+1}{{\sigma}_{-}}^{i} )]
\eea
${\Delta}_{i,i+1}$ is the constant electrostatic coupling amplitude. The
tunable parameters of the system: ${{B_x}^{1,N}}$ the Josephson coupling
of the leftmost and rightmost Squids. ${{B_z}^{i}}$ is the electrostatic 
potential of the island and $J_{i,i+1}$ is the Josephson coupling between
the neighboring island. We are interested in the charge degeneracy point
at this point the most favorable state of the system is the antiferromagnetic
configuration $(|010101.....>$ and $|101010.....>)$. 
They start with one of the antiferromagnetic states and transfer
the charge of every island to the right by two sites to achieve pumping.
They have implemented it by applying identical pulse sequence to every second
island \cite{ravi}.  So we can write ${{B_x}^{1}}= J_{2,3}=J_{4,5}= J_{e,o}$ and
${{B_x}^{N}}= J_{1,2}=J_{3,4}= J_{o,e}$. $J_{e,o}$ and $J_{o,e}$ are 
respectively the even-odd and odd-even Josephson couplings. They have also
considered 
a difference between the charging energies 
between the odd and even sites. In our theoretical analysis, we consider
the Josephson couplings for even  and odd sites as respectively  
$E_{J_1}  ~=~ E_{J_0} ( 1 - {{\delta}_1} (t) )$ and 
$E_{J_1}  ~=~ E_{J_0} ( 1 + {{\delta}_1} (t) )$. The charging energies of even
and odd sites are respectively ${B_{z1}}  ~=~ {B_0} ( 1 - {{\delta}_2} (t) )$    
and ${B_{z2}}  ~=~ {B_0} ( 1 + {{\delta}_2} (t) )$. 
One can write the Hamiltonian in terms of spin operators:
\bea
H & = & - \sum_{n} {E_{J_0}} ( 1 - (-1)^{n} {{\delta}_1} (t) ) 
( {{S}_{+}}^{n}{{S}_{-}}^{n+1} +
{{S}_{+}}^{n+1}{{S}_{-}}^{n} ) \nonumber\\
& &  + \sum_{n} {\Delta} {{S}_z}^{n}{{S}_z}^{n+1} 
- \frac{1}{2} \sum_{n} {B_0} ( 1 - (-1)^{n} {{\delta}_2} (t) ) {{S}_z}^{n} 
\eea
One can express 
spin chain systems to a spinless fermion systems through 
the application of Jordan-Wigner transformation. In Jordan-Wigner transformation
the relation between the spin and the electron creation and annihilation operators
are  
$ S_n^z  =  \psi_n^{\dagger} \psi_n - 1/2 ~$, 
$ S_n^-  =   \psi_n ~\exp [i \pi \sum_{j=-\infty}^{n-1} n_j]~$, 
$ S_n^+  =  \psi_n^{\dagger} ~\exp [-i \pi \sum_{j=-\infty}^{n-1} n_j]~$,
\cite{gia2}, where $n_j = \psi_j^{\dagger} \psi_j$ is the fermion number at site $j$. 
\bea
{H}~&=& - \frac{E_{J0}}{2} ~\sum_n ~ ( 1 - (-1)^{n} {{\delta}_1}(t) )
(\psi_{n+1}^{\dagger} \psi_n + 
\psi_n^{\dagger} \psi_{n+1}) \nonum \\
&& + \Delta \sum_n (\psi_n^{\dagger} \psi_n - 1/2) (\psi_{n+1}^{\dagger} 
\psi_{n+1} - 1/2) ~,\nonum \\
&&  - \frac{B_0}{2} \sum_{n} ( 1 - (-1)^{n} {{\delta}_2}(t) )     
(\psi_n^{\dagger} \psi_n - 1/2).
\label{ha1}
\eea
Our approach is completely analytical. 
Before we proceed further for continuum field theoretical 
study of these model Hamiltonians, we would like to explain the
basic aspects of quantum Cooper pair pumping in terms of 
spin pumping physics of our model Hamiltonians: 
An adiabatic sliding motion of one dimensional potential,
in gapped Fermi surface (insulating state), pumps an integer numbers
of particle 
per cycle.
In our case the transport of Jordan-Wigner
fermions (spinless  fermions) is nothing but the transport of spin from one end
of the chain to the other end because the number operator of spinless fermions
is related with the z-component of spin density \cite{cal}. 
We shall see that non-zero ${\delta} (t)$ introduces the gap at 
around the
Fermi point and the system is in the insulating state (Peierls insulator).
In this phase spinless fermions form the bonding orbital between the
neighboring sites, which yields a valance band in the momentum space.
It is well known that the physical behavior of the system is identical
at these two Fermi points. 
One can analyses this double
degeneracy point, 
from the seminal paper of Berry \cite{berry}, 
It appears as source and sink vector fields defined 
in the generalized
crystal momentum space.
${B_n} (K) = {{\nabla}_K} \times {A_n} (K)$, and 
${A_n} (K) = \frac{i}{2 \pi} <n (K)| {{\nabla}_K} | n(K)>$, 
where $K = (k, {\delta} (t) )$.
Here $B_n$ and $A_n$ are the fictitious magnetic field (flux) and 
vector potential of the
nth Bloch band respectively.
The degenerate points behave as a magnetic monopole in the
generalized momentum space \cite{berry}, whose magnetic unit can be shown to be
$1$, analytically \cite{shin,berry}
\beq
\int_{S1} ~dS \cdot B_{\pm} ~=~ \pm 1
\eeq 
positive and negative signs of the above equations are respectively
for the conduction and valance band
meet at the degeneracy points.
$S_1 $ represent an arbitrary closed surface which enclose the
degeneracy point.
In the adiabatic process the parameter ${\delta} (t)$ is changed
along a loop ($\Gamma$) enclosing the origin (minima of the system).
We define the expression for spin current ($I$) from the analysis
of Berry phase.
Then according to the original
idea of quantum adiabatic particle transport \cite{thou1,thou2,shin,avron},
the total number of spinless fermions ($I$)
which are transported from one side of this system to the other is equal to the
total flux of the valance band, which penetrates the 2D closed sphere
($S_2 $) spanned by
the $\Gamma$ and Brillioun zone \cite{shin}.
\beq
 I = \int_{S_2} dS \cdot B_{+1} ~=1
\eeq
We have already understood that quantized
spinless fermion transport is equivalent to the spin transport \cite{cal}.
We will interpret this equation more physically after Eq. 18 .
This quantization is
topologically protected against the other perturbation as long as 
the gap along the
loop remains finite \cite{shin,avron}.
Studies of spin pumping 
explain the
stabilization of quantized spin pumping against z-component of 
exchange interactions.\\

We recast the spinless
fermions operators in terms of field operators by this relation 
\beq
{\psi}(x)~=~~[e^{i k_F x} ~ {\psi}_{R}(x)~+~e^{-i k_F x} ~ {\psi}_{L}(x)]
\eeq
where ${\psi}_{R} (x)$ and ${\psi}_{L}(x) $ describe 
the second-quantized fields of right- and 
left-moving fermions respectively.
We want to express the fermionic fields in terms of bosonic field by this relation 
\beq
{{\psi}_{r}} (x)~=~~\frac{U_r}{\sqrt{2 \pi \alpha}}~~e^{-i ~(r \phi (x)~-~ \theta (x))} 
\eeq
$r$ is denoting the chirality of the fermionic fields,
 right (1) or left movers (-1).
The operators $U_r$ are operators that commute with the bosonic field. $U_r$ of different species
commute and $U_r$ of the same species anticommute. $\phi$ field corresponds to the 
quantum fluctuations (bosonic) of spin and $\theta$ is the dual field of $\phi$. 
They are
related by this relation $ {\phi}_{R}~=~~ \theta ~-~ \phi$ and  $ {\phi}_{L}~=~~ \theta ~+~ \phi$.

Using the standard machinery of continuum field theory \cite{gia2}, 
we finally get the bosonized Hamiltonians
as 
\bea
H_{0}~&=&~v_0 \int_{o}^{L} \frac{dx}{2 \pi} \{ {\pi}^2 : {\Pi}^2 : ~+~ :[ {\partial}_{x} \phi (x) ]^2 : \}  \nonum\\
&&~+~ \frac{2 \Delta}{{\pi}^2 }\int ~ dx~:[ {\partial}_{x} {{\phi}_L} (x) ]^2  :
+ :[ {\partial}_{x} {{\phi}_R} (x) ]^2 : \nonum\\
&&~+~ \frac{4 \Delta}{{\pi}^2 }\int ~ dx~ 
 ({\partial}_{x} {{\phi}_L} (x) ) ({\partial}_{x} {{\phi}_R} (x) ) 
\label{bos1}
\eea
$H_{0}$ is the gapless Tomonoga-Luttinger liquid part of the Hamiltonian
with $v_0~=sink_F$. 
\\
After continuum field-theory the Hamiltonian become
\bea
H & = & {H_0} 
+ \frac{{E_{J_0}} {{\delta}_1}(t) }{2 {{\pi}^2} {\alpha}^2} 
\int dx : cos[2 \sqrt{K} {\phi} (x)]: \nonum\\ 
& & + \frac{{B_0} {{\delta}_2}(t) }{2 {\pi} {\alpha}} 
\int dx : cos[2 \sqrt{K} {\phi} (x)]: \nonum\\
& & + \frac{\Delta}{2 {{\pi}^2} {\alpha}^2} \int dx 
: cos[4 \sqrt{K} {\phi} (x)]: 
- \frac{B_0}{2} \int dx {{\partial}_x} {\phi}(x)
\eea 
The second term of the Hamiltonian for NN exchange interaction has
originated from the XY interaction. This dimerization is the
spontaneous dimerization, 
i.e., the infinitesimal variation of 
Josephson coupling in lattice sites, is sufficient to produce a 
gap around the
Fermi points.
The third term of the Hamiltonian arises due to the site dependent
on-site charging energies modulated by the gate voltage. It yields the 
staggered phase of the
system. The effect of applied gate voltage on the Cooper pair box appears
as an effective magnetic field and also as a staggered magnetic in the
spin representation of the model Hamiltonian.
The system is in the mixed phase when both interactions 
(2nd and 3rd terms of the Hamiltonian) are in equal
magnitude otherwise the system is in any one of the states of the
mixed phase depending on the strength of the couplings. The last term
can be absorbed in the Hamiltonian through the proper shifting of
the wave function. 
So when ${1/2} < K < 1$ only these (2nd and 3rd terms of the Hamiltonian)
time dependent dimerizing field
(third term of Eq. 17)
is relevant and lock the phase operator at ${\phi} = 0 + \frac{n \pi}{\sqrt{K}}$.
Now the locking potential slides adiabatically (here the cyclic
magnetic flux and gate voltage 
fields that produces the dimerization). Speed of the sliding potential is low
enough such that system stays in the same valley, i.e., 
there is no scope to jump
onto the other valley.  
The system will acquire $2 \pi$ phase during one
complete cycle of  
dimerizing field. 
This is the basic mechanism of Cooper pair (spin pumping) of
our system. 
The quantized Cooper pair transport of this scenario can be generalized up to the 
value of $\Delta$  for which 
$K$ is greater than 1/2  
. In this limit, z-component
of exchange interaction  
has no effect on the Cooper pair pumping
physics of  
Hamiltonian.  
This 
expection is easily verified when we notice the physical meaning of the phase
operator ($\phi$ (x)). Since the spatial derivative of the phase operator 
corresponds to the z-component of spin density (Cooper pair density), 
this phase operator is
nothing but the minus of the spatial polarization of the z-component of
spin, i.e., 
$ P_{s^z}~= - \frac{1}{N} \sum_{j=1}^{N} j {S_j}^z $. Shindou
has shown explicitly 
the equivalence between these two consideration \cite{shin}. During the
adiabatic process $ < {\phi}_{t} >$ changes monotonically and acquires
- $2 \pi$ phase. In this process $ {P_s}^{z} $ increases by 1 per cycle.
We define it analytically as
\beq
{\delta} {P_s}^{z} = \int_{\Gamma} d {P_s}^{z}
= - \frac{1}{2 \pi} \int dx {{\partial}_x} <{\phi} (x) > = 1
\eeq
This physics always hold as far as the system is locked by the sliding 
potential and ${\Delta} < 1$ \cite{shin}. 
The above equation for Cooper pair  transport is physically consistent
with the Eq. 13 (based on Berry phase analysis) of Cooper pair current.
The quantized Cooper pair transport of this scenario can be generalized up to the
value of $\Delta$  for which
$K$ is greater than 1/2
. In this limit, z-component
of exchange interaction 
has no effect on the Cooper pair pumping
physics of our system.\\  
In summary, we have presented the theoretical explanation of adiabatic Cooper pair 
pumping of experimentally reliazable stabilized charge pumping
scheme of an array of Cooper pair boxes. The charge state of our system is
definite.\\
The author (SS) would like to acknowledge The Center for Condensed Matter 
Theory of IISc and also
Mr. Vasudeva for reading the manuscript very critically.

\end{document}